\pdfoutput=1
\documentclass[conference]{IEEEtran}

\ifCLASSOPTIONcompsoc
  \usepackage[nocompress]{cite}
\else
  \usepackage{cite}
\fi
\ifCLASSINFOpdf
\else
\fi

\usepackage{amsmath}
\usepackage{algorithmic}

\usepackage{subfigure}
\usepackage{graphicx}
\usepackage{multirow}
\usepackage{algorithm}
\usepackage{color}
\usepackage{amsmath}
\usepackage{mathabx}
\usepackage[hyphens]{url}
\usepackage{ctable}

\newcommand{\squishlist}{
    \begin{list}{$\bullet$}
    { \setlength{\itemsep}{0pt}
        \setlength{\parsep}{3pt}
        \setlength{\topsep}{3pt}
        \setlength{\partopsep}{0pt}
        \setlength{\leftmargin}{1.5em}
        \setlength{\labelwidth}{1em}
        \setlength{\labelsep}{0.5em} } }

\newcommand{\squishlisttwo}{
    \begin{list}{$\bullet$}
        { \setlength{\itemsep}{0pt}
            \setlength{\parsep}{0pt}
            \setlength{\topsep}{0pt}
            \setlength{\partopsep}{0pt}
            \setlength{\leftmargin}{2em}
            \setlength{\labelwidth}{1.5em}
            \setlength{\labelsep}{0.5em} } }

\newcommand{\squishend}{
    \end{list}  }

\usepackage{fancyhdr}
\usepackage{kantlipsum}

\fancypagestyle{plain}{
\fancyhf{}
\fancyhead[C]{Conference on \LaTeX} 

}
\usepackage{eso-pic}

\begin{document}

\AddToShipoutPictureBG*{%
\AtPageUpperLeft{%
\setlength\unitlength{1in}%
\hspace*{\dimexpr0.5\paperwidth\relax}
\makebox(0,-0.75)[c]{\textbf{2016 IEEE/ACM International Conference on Advances in Social Networks Analysis and Mining (ASONAM)}}}}

\IEEEoverridecommandlockouts
\IEEEpubid{\parbox{\columnwidth}{\textbf{IEEE/ACM ASONAM 2016, August 18-21, 2016, San Francisco, CA, USA \\978-1-5090-2846-7/16/\$31.00 \copyright2016 IEEE} \hfill}
\hspace{0.7\columnsep}
\makebox[\columnwidth]{}}

%
\title{Understanding Citizen Reactions and Ebola-Related Information Propagation on Social Media}

\author{\IEEEauthorblockN{Thanh Tran}
\IEEEauthorblockA{Department of Computer Science\\
Utah State University, Logan, UT 84322\\
Email: thanh.tran@aggiemail.usu.edu}
\and
\IEEEauthorblockN{Kyumin Lee}
\IEEEauthorblockA{Department of Computer Science\\
Utah State University, Logan, UT 84322\\
Email: kyumin.lee@usu.edu}
}


%


\maketitle

\begin{abstract}
In severe outbreaks such as Ebola, bird flu and SARS, people share news, and their thoughts and responses regarding the outbreaks on social media. Understanding how people perceive the severe outbreaks, what their responses are, and what factors affect these responses become important. In this paper, we conduct a comprehensive study of understanding and mining the spread of Ebola-related information on social media. In particular, we (i) conduct a large-scale data-driven analysis of geotagged social media messages to understand citizen reactions regarding Ebola; (ii) build information propagation models which measure locality of information; and (iii) analyze spatial, temporal and social properties of Ebola-related information. Our work provides new insights into Ebola outbreak by understanding citizen reactions and topic-based information propagation, as well as providing a foundation for analysis and response of future public health crises.
\end{abstract}


\section{Introduction}
\vspace{-5pt}



The World Health Organization declared that Ebola became an international public health problem in August 8, 2014. A person infected by Ebola entered the US in September 19, 2014 and passed away on October 8, 2014. Two nurses, who cared the infected person, also got infected. A physician was positively diagnosed on October 23, 2014.

Ebola outbreak has created not only severe public health challenges directly in the areas of infections, but also a tumultuous public health information crisis in the US and abroad. For example, when the 2014 Ebola crisis in West Africa erupted, there were strong US-based reactions with schools closed, voluntary quarantines of students who had merely visited a country in Africa, HazMat teams deployed to scenes of everyday illness, among other potentially costly reactions.

Ebola outbreak caused public panic and concerns, and people expressed their responses on social media. Examples of tweets expressing concerns about Ebola on Twitter are ``What if we all put In so much hard work \& when we graduate \& get our diplomas, we die of Ebola'', and ``Another \#Ebola patient in \#Orlando? Just saw police escorted ambulance going to \#MCO airport''. While the sharing of public health information and personal health updates (e.g., ``I'm sick'') has grown with the commensurate spread of social media \cite{ 
rodriguez2015makes,fung2014ebola,oyeyemi2014ebola}, there is a research gap in our understanding of the landscape of reactions to severe outbreaks, the factors affecting these reactions, and the spatial and temporal patterns of these reactions.

In this paper, we aim to (i) conduct a data-driven analysis of the Ebola outbreak to understand citizen reactions, (ii) build novel Ebola-related information propagation models, and (iii) examine how spatial, temporal and social properties are related to Ebola-related information propagation. Specifically, we seek to answer following research questions: What kind of Ebola related messages people posted? Can we group their messages by topics? 
Can we measure locality of Ebola-related information? By conducting two US Ebola case studies, can we understand how Ebola-related information was propagated over time? Which topic of citizen responses was more broadly propagated?

To answer these questions, we collected 2 billion tweets from Twitter and extracted geotagged Ebola-related tweets. Then, we make the following contributions in this paper:

\squishlist
\item We conducted a comprehensive data-driven analysis of the geotagged Ebola-related tweets to understand 
language distribution and geographic distribution. 
\item We built topic models from citizen reactions and found six topics. Then we measured how locally or globally each topic was spread.
\item Finally, we defined three locality measures called Ebola focus, entropy and spread. Then, we conducted two US case studies like Dallas case and New York City case with analyzing spatial, temporal and social properties.
\squishend

\begin{figure*}
	\centering
	\includegraphics[width=0.70\linewidth]{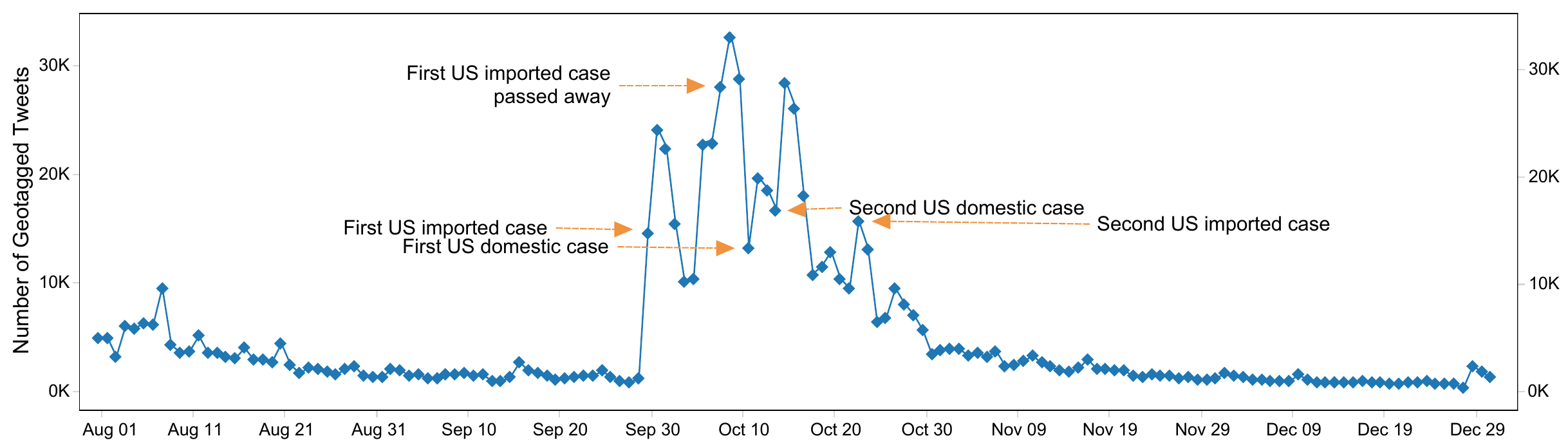}
	\caption{A distribution of the number of geotagged tweets over time.}
	\label{fig:TweetFreq}
	\vspace{-15pt}
\end{figure*}

\section{Related Work}
\vspace{-5pt}

Ebola virus became a very serious problem in the world, researchers began studying Ebola-related information on social media. Seltzer et al. \cite{seltzer2015content} collected 1,217 Ebola-related images posted on Instagram and Flickr, and grouped the images by 9 themes. Lazard et al. \cite{lazard2015detecting} collected 2,155 tweets containing a hashtag \#CDCchat, which were posted from the Centers for Disease Control and some Twitter users, and then found 8 topics from the tweets. Fung et al. \cite{fung2014ebola} analyzed how the frequency of Ebola-related tweets is correlated with the frequency of searches on Google. Alicino et al. \cite{alicino2015assessing} analyzed Google search queries related to the Ebola outbreak to understand the correlation between the number of web searches and the number of Ebola cases.

Compared to the previous research works, we collect and analyze a large-scale Ebola-related tweets, especially focusing on geotagged tweets. We build topic models to understand what kind of responses people had. 
Then, we analyze how citizen responses are related to spatial, temporal and social properties by designing Ebola propagation models. 

\section{Dataset}
\vspace{-5pt}

Ebola virus disease/Ebola was first identified in 1976 in Nzara, South Sudan\footnote{https://en.wikipedia.org/wiki/Ebola\_virus\_disease}. Between 1976 and 2013, 1,716 people got infected by Ebola. According to the report of Centers for Disease Control and Prevention (CDC), there were 20,171 Ebola cases and 7,889 patients were died in 2014\footnote{http://www.cdc.gov/vhf/ebola/outbreaks/2014-west-africa/previous-case-counts.html}. In particular, the total number of cases in 2014 increased from 1,437 to 20,171 between August 2014 and December 2014.

In this paper, we focus on Ebola-related tweets posted between August 2014 and December 2014 (i.e., during the major Ebola outbreak). First, we collected 2 billion tweets by using Twitter Streaming APIs and running two crawlers -- one collecting random tweets (\emph{random crawler}) and the other one collecting only geotagged tweets (\emph{geo crawler}). Out of 2 billion tweets, we searched a keyword ``ebola'' in each of 90 languages to extract tweets containing ``ebola''. Then, we grouped the retrieved tweets by a language, removed duplicated tweets, and counted the number of tweets in each group. We removed if a group consists of less than 100 tweets (i.e., the language was used in less than 100 tweets). Out of 90 languages, each of 35 ones was used in at least 100 tweets. The total number of the remaining Ebola-related tweets was 1,167,539 tweets (569,888 geotagged tweets from \emph{geo crawler} and 597,651 non-geotagged tweets from \emph{random crawler}).

To estimate geolocation of the remaining 597,651 non-geotagged tweets, we estimated a home location of the non-geotagged tweet posters/users. Based on the previous works estimating a Twitter user's home location \cite{cheng2010you, davis2011inferring}, we estimated a non-geotagged tweet poster's home location as follows: first, we collected 200 most recent tweets from each user. Among the 200 tweets, we checked how many tweets contained geolocation information. If at least two out of 200 tweets of a user contained geolocation information, we would consider estimating a home location of the user. Otherwise, we filtered the user's Ebola-related tweets from the home location estimation process. Given a user's geotagged tweets, we measured the median latitude and median longitude, and used them as geolocation of the user's Ebola-related tweets. Finally, we estimated geolocation of 199,576 non-geotagged tweets. In the following sections, we use 769,464 (569,888+199,576) geotagged tweets\footnote{From now on, we call both geotagged tweets and geolocation-estimated tweets as geotagged tweets.}, posted by 422,596 users for in-depth analysis and study.

\section{Analysis of All Geotagged Tweets}
\vspace{-5pt}


In this section, we analyze the 769,464 geotagged tweets
. First, we analyzed a distribution of the number of geotagged tweets in our dataset. Figure~\ref{fig:TweetFreq} shows how many geotagged tweets were posted over time. We clearly observed that most tweets were posted between September 30 and October 30, 2014. We highlighted dates of the US imported and domestic cases in the figure. According to a Wikipedia page \cite{ebolainfo:Misc}, the first US imported case happened when Thomas Eric Duncan visited his relatives in Dallas, Texas traveled from Liberia on September 30, 2014. Then, he died on October 8, 2014 and infected two nurses (two US domestic cases) who took care of him at Texas Health Presbyterian Hospital. The second US imported case happened when a physician Craig Spencer from Guinea was positively diagnosed on October 23, 2014 in New York City. 



Next, we analyzed what languages users used in Ebola-related tweets. English 
was used the most, taking 62.24\% of tweets. Then, Spanish, Portuguese, French, Japanese, Italian were used in 18.63\%, 8.18\%, 3.07\%, 1.16\% and 1.14\% of tweets, respectively. 5.58\% of tweets were posted in other languages.


An interesting question is ``From where tweets were posted?''. To answer the question, we analyzed geolocation information associated with tweets in our dataset. In particular, we grouped tweets by a country and counted the frequency. Top 5 countries were United States, Spain, Brazil, U.K. and Niger where 47.06\%, 8.03\%, 6.72\%, 5.24\%, 3.12\% of tweets were posted from, respectively. Interestingly, most top countries were the ones which had evacuated or imported cases. The remaining top countries were their neighbor countries. 

\begin{table*}

    \centering
    \caption{6 LDA topics, number of associated tweets and sample tweets.}
    \scriptsize

	\begin{tabular}	{p{1.3in}p{5.5in}}
		\toprule
		Topics ($|$tweets$|$)& Sample Tweets \\
		\midrule
		1. Ebola cases in US (33,828)   &
a. Ebola infected missionaries returning to US for treatment. NYC hospitals on alert but no imminent threat. @ABC7NY \#ABC7WakeUp  \newline
b. Top reason Houston is better than Dallas RT @KUT: A Dallas hospital says it has isolated a patient who may have contracted the \#Ebola virus \\
		\midrule 2. Ebola outbreak in the world (37,195)	
& a. The World Health Organization says the Ebola outbreak in West Africa is spreading faster than efforts to control it. \#GNreport \newline
 b. Nigeria Government will be like.....we are fighting EBola; but doctors are on strike for months now! @Emmydny\\
        \midrule 3. Fear and pray (44,307) &
a. Praying God will catch this strain of Ebola. Jesus be a mighty fence! \newline
b. This EBOLA scares me.... I'm preparing for a break out\\
        \midrule 4. Ebola spread and warning (44,301) &
a. The United States has issued travel warning for three West African countries hit by the deadly Ebola virus outbreak. \#GNreport \newline
b. With this Ebola crisis going on, why aren't we stopping flights in and out of Africa. It seems very dangerous to allow people to travel. \\
        \midrule 5. Jokes, swear and disapproval of joke (63,648) &
a. Ebola jokes are for stupid people...bananyana! very unfunny \newline
b. ``@PrncessAriel: Ebola is scaring the shit out of me'' right? Shit's no joke \\
		\midrule 6. Impact of Ebola to daily life (78,444) &
a. Feeling not to work tomorrow cos there's bare fobs and I'm shook of Ebola \newline
b. Talk on \#Ebola: \#Ebola stopped me from going to school. \#Ebola stopped me from playing.\#Ebola stopped me visiting my friends @PlanGlobal \\
        \bottomrule
	\end{tabular}
	\label{table:LDATopicTweetExample}
	\vspace{-15pt}
\end{table*}

\section{Analysis of Geotagged English Tweets}
\vspace{-5pt}
So far, we analyzed all the geotagged tweets. Now we turn to analyze geotagged English tweets by topic modeling since English is a major language (62.24\% of tweets) in our dataset. In particular, we are interested in understanding what kind of topics users discussed.

To answer the question, first we extracted geotagged English tweets from our dataset. Then, we removed news tweets each of which contained a title of a news article (e.g., CNN or New York Times articles) because these tweets did not contain a user's opinion or response. Finally, 301,723 geotagged English tweets were extracted and used in this section.

\label{sec:topic}

Next, we developed topic models based on Latent Dirichlet Allocation (LDA) \cite{blei2003latent}. 
Given a collection of Ebola-related messages/tweets, output of LDA is a group of topics each of which contains a set of messages belonging to the topic. From each topic, we can analyze what kind of Ebola-related messages/tweets people posted to social media toward understanding their responses.

\begin{figure}
	\centering
	\includegraphics[width=0.43\textwidth]{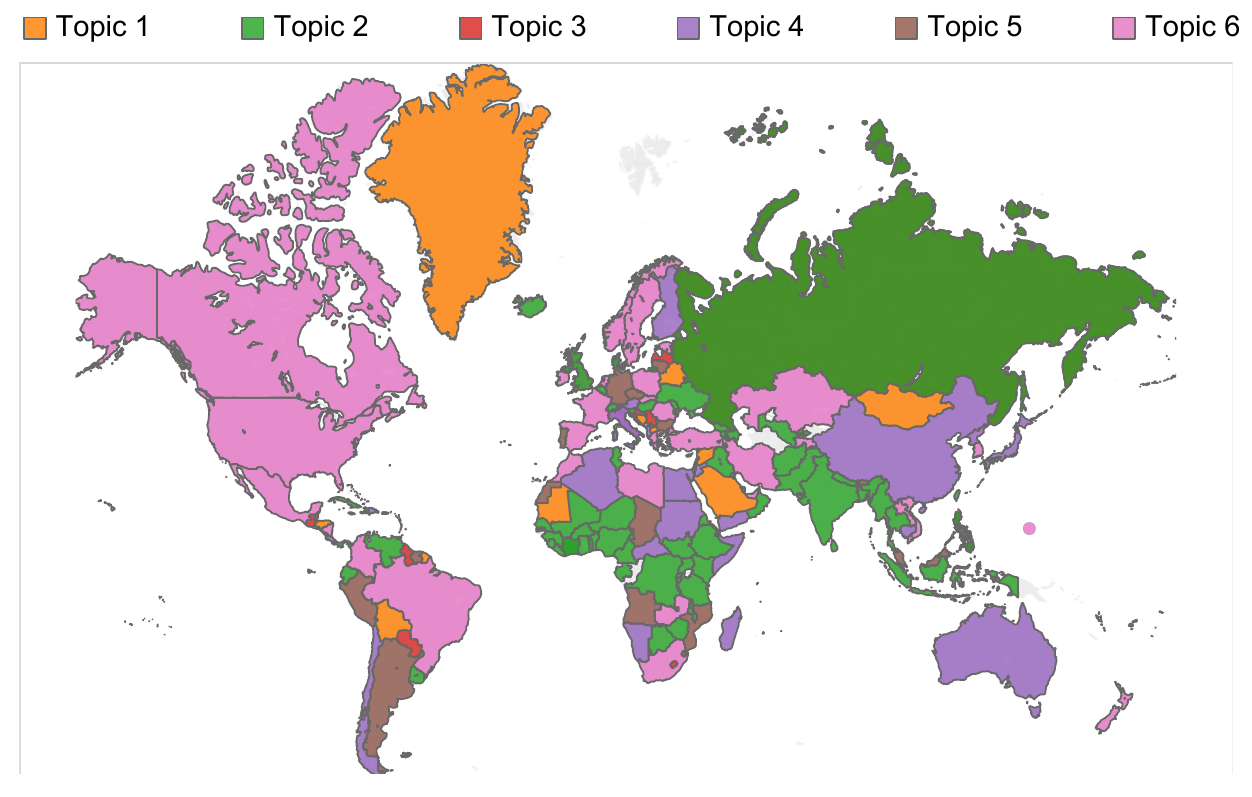}
	\caption{Major LDA topics in each country.}
	\label{fig:LDATopicWorldMap}
	\vspace{-15pt}
\end{figure}

Since LDA works better with long documents, instead of using each tweet/message as a document, we aggregated each user's Ebola-related tweets to make a longer document. Based on a perplexity score, we chose \emph{K} as 6. In other words, we found six topics from the collection of Ebola-related tweets. Table \ref{table:LDATopicTweetExample} presents inferred topic labels, the number of associated tweets and sample tweets. Topic 1 and 2 are related to ``Ebola cases in US'' and ``Ebola outbreak in the world'', respectively. Topic 3 contains tweets related to ``Fear and pray''. Users in topic 4 mentioned about ``Ebola spread and warning''. Topic 5 represents ``Jokes, swear and disapproval of joke''. Topic 6 contains tweets about ``Impact of Ebola to daily life'' such as working and schooling. Among the six topics, topics 5 and 6 contained the largest number of tweets while topic 1 contained the smallest number of tweets. In other words, many Twitter users talked about jokes, swear and disapproval of joke, and impact of Ebola to their daily life.  Compared to 8 topics found by analyzing Centers for Disease Control and Prevention's Ebola live Twitter chat at \cite{lazard2015detecting},  our topics are more general. 

Next, we visualized which topic was the most popular in each country on the world map as shown in Figure~\ref{fig:LDATopicWorldMap}. Topic 2 ``Ebola outbreak in the world'' was the most popular topic in the world. Specially it was popular in most Ebola countries like Guinea, Sierra Leone, Liberia, Nigeria, Senegal, U.K. and Mali. The next most popular topic was topic 6 ``Impact of Ebola to daily life'', which was mostly talked in Ebola countries like US and Spain, and some European countries nearby Ebola countries (e.g., France, Morocco, Ireland, Poland and Netherlands nearby Spain, U.K. and Italy).

\section{Mining Ebola-Related Information Propagation}
\vspace{-5pt}
\label{PropagationAnalysis}

In this section, we analyze Ebola-related information propagation on social media. To measure how Ebola-related information was spread, we conducted two US imported case studies. According to a Wikipedia page \cite{ebolainfo:Misc}, the first US imported case happened on September 30, 2014 in Dallas. The second US imported case happened on October 23, 2014 in New York City. These two imported cases became two events/sources of Ebola-related information to measure how such Ebola-related information was spread. We chose \emph{Texas Health Presbyterian Hospital} at Dallas and \emph{Bellevue} at New York City as two center points where Ebola-related tweets first appeared and began spreading to other regions. Note that we used geotagged tweets posted in US in this section since we focused on two US imported cases.

\begin{figure}
	\centering
	\subfigure[Dallas case] 
	{
		\label{fig:DallasDailyDis}
		\includegraphics[width=0.40\linewidth, height=0.9in]{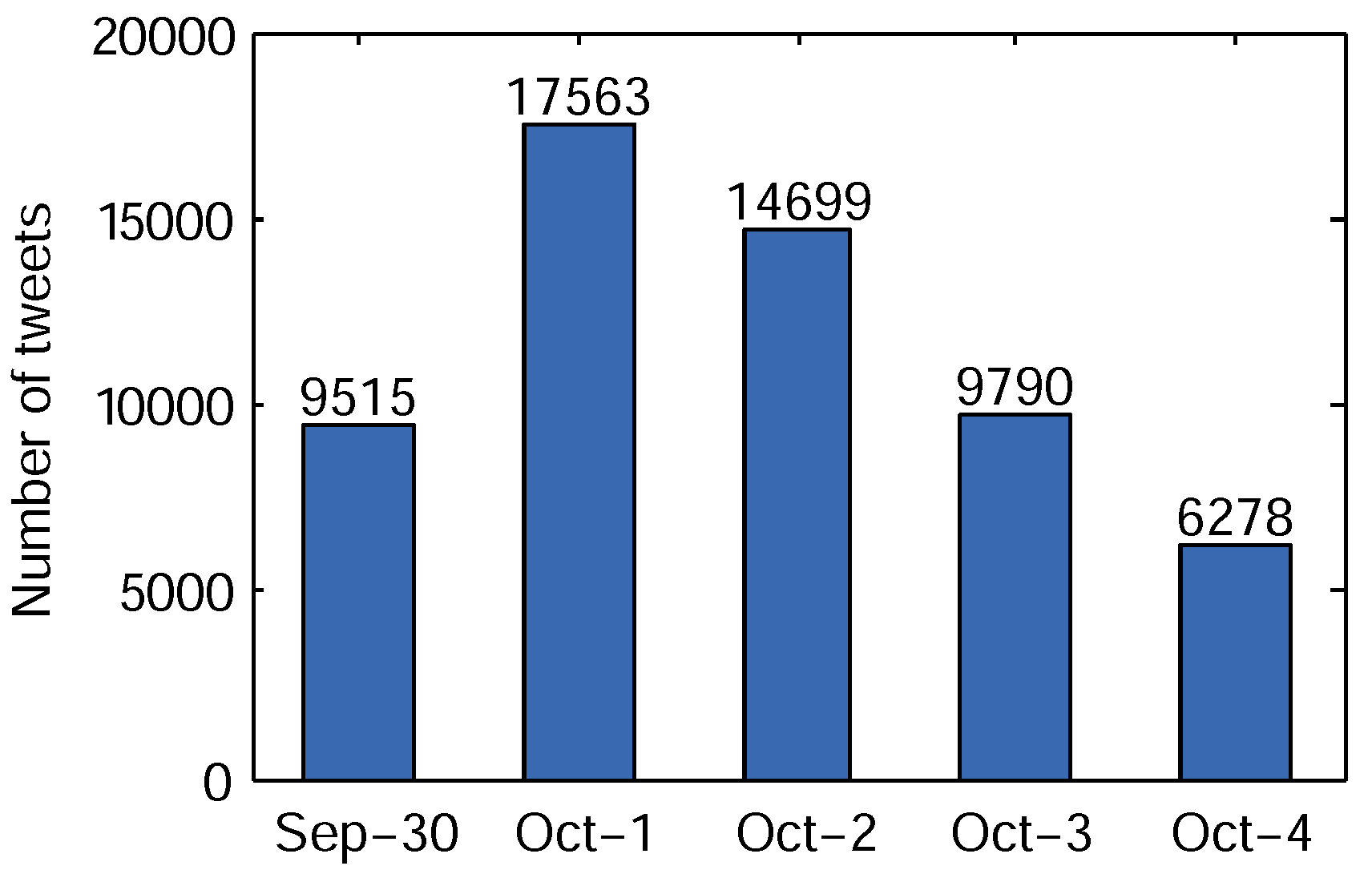}
	}
	\subfigure[New York City case] 
	{
		\label{fig:NewYorkDailyDis}
		\includegraphics[width=0.40\linewidth,height=0.9in]{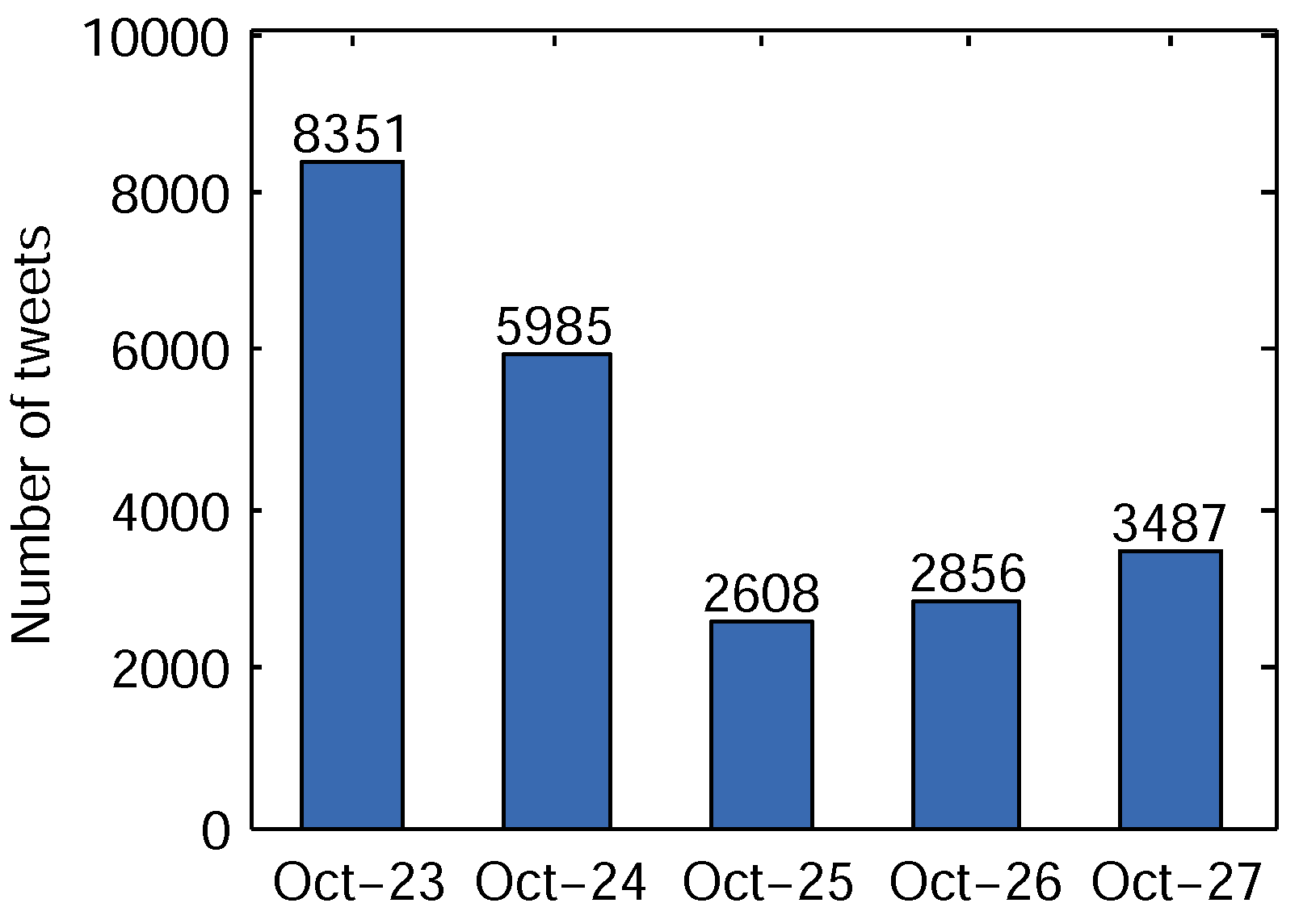}
	}
	\caption{The number of Ebola-related tweets posted in US for five days which were related to two cases.}
	\label{fig:DailyDisDallasNewYork}
	\vspace{-10pt}
\end{figure}

Figure~\ref{fig:DailyDisDallasNewYork} shows the number of Ebola-related tweets of each case during five days since each case was confirmed (i.e., positively diagnosed). Interestingly, even though the first case was confirmed on September 30, a peak day was October 1 while a peak day of the second case was October 23, the same day, when the case was confirmed. We conjecture that it took time for social media users in US to pay attention on the first case (i.e., 1 day delay), but they responded so quickly when the second case happened.

Authors of the previous works \cite{crane2008robust} suggested that a peak day is the starting day to analyze spread of information, so we chose October 1, 2014 and October 23, 2014 as two starting days to measure the propagation of Ebola-related tweets. Kwak et al. \cite{kwak2010twitter} found that the half of tweets were propagated within an hour, and 75\% of tweets were propagated within a day for each event. Based on the results, we extracted 32,262 tweets posted in US between October 1 and October 2, 2014 for the first case and 14,336 tweets posted in US between October 23 and October 24, 2014 for the second case. Overall we extracted 46,598 tweets posted in US from our dataset to investigate how spatial, temporal and social properties were related to information propagation.

Before performing our analysis, we first define how to measure locality of each case/event.

\vspace{-5pt}
\subsection{Locality Measures}
\vspace{-5pt}

\begin{figure}
	\centering
	\subfigure[Dallas case] 
	{
		\label{fig:SpatialCDFDallas}
		\includegraphics[width=0.40\linewidth]{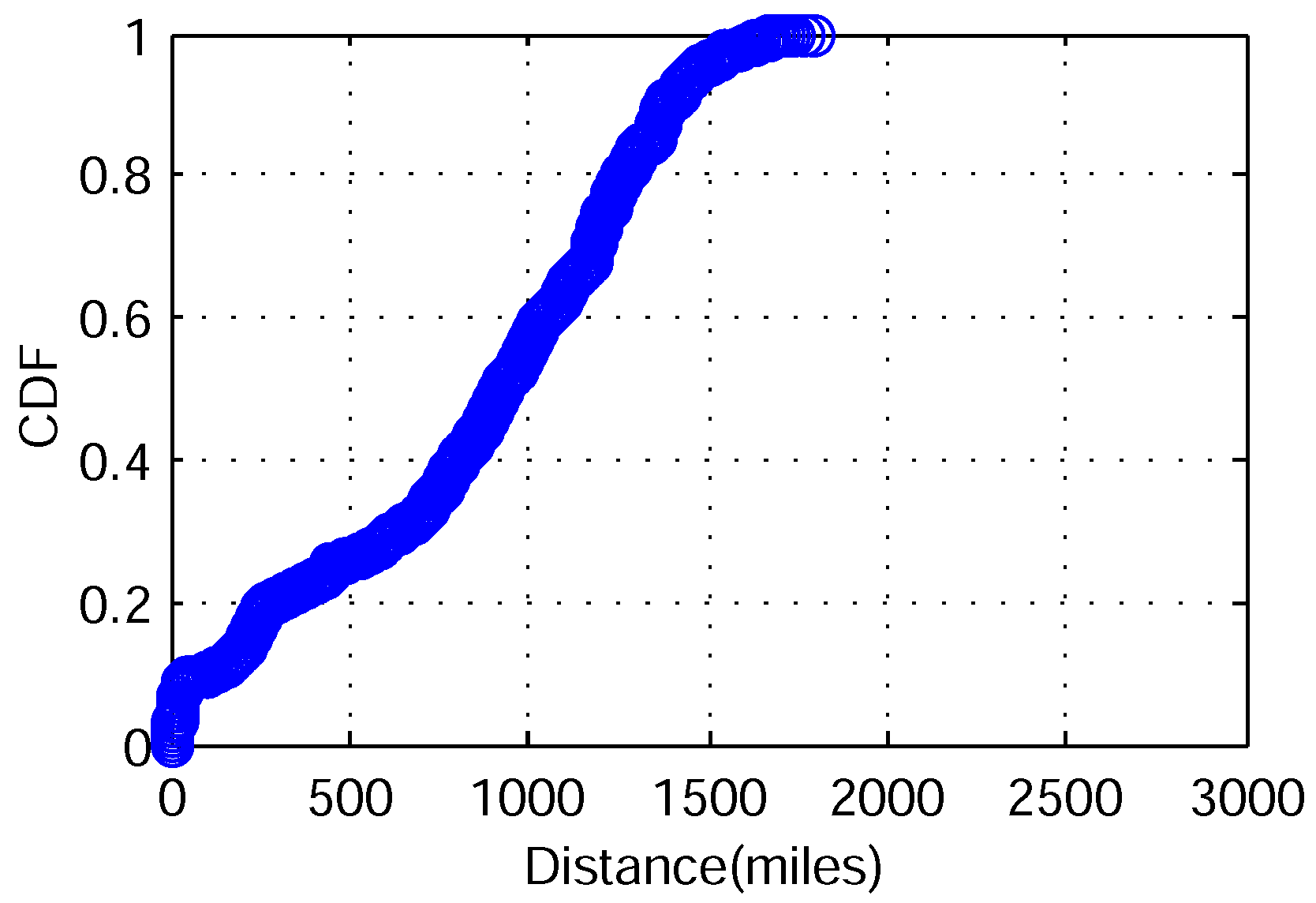}
	}
	\subfigure[New York City case] 
	{
		\label{fig:SpatialCDFNewYork}
		\includegraphics[width=0.40\linewidth]{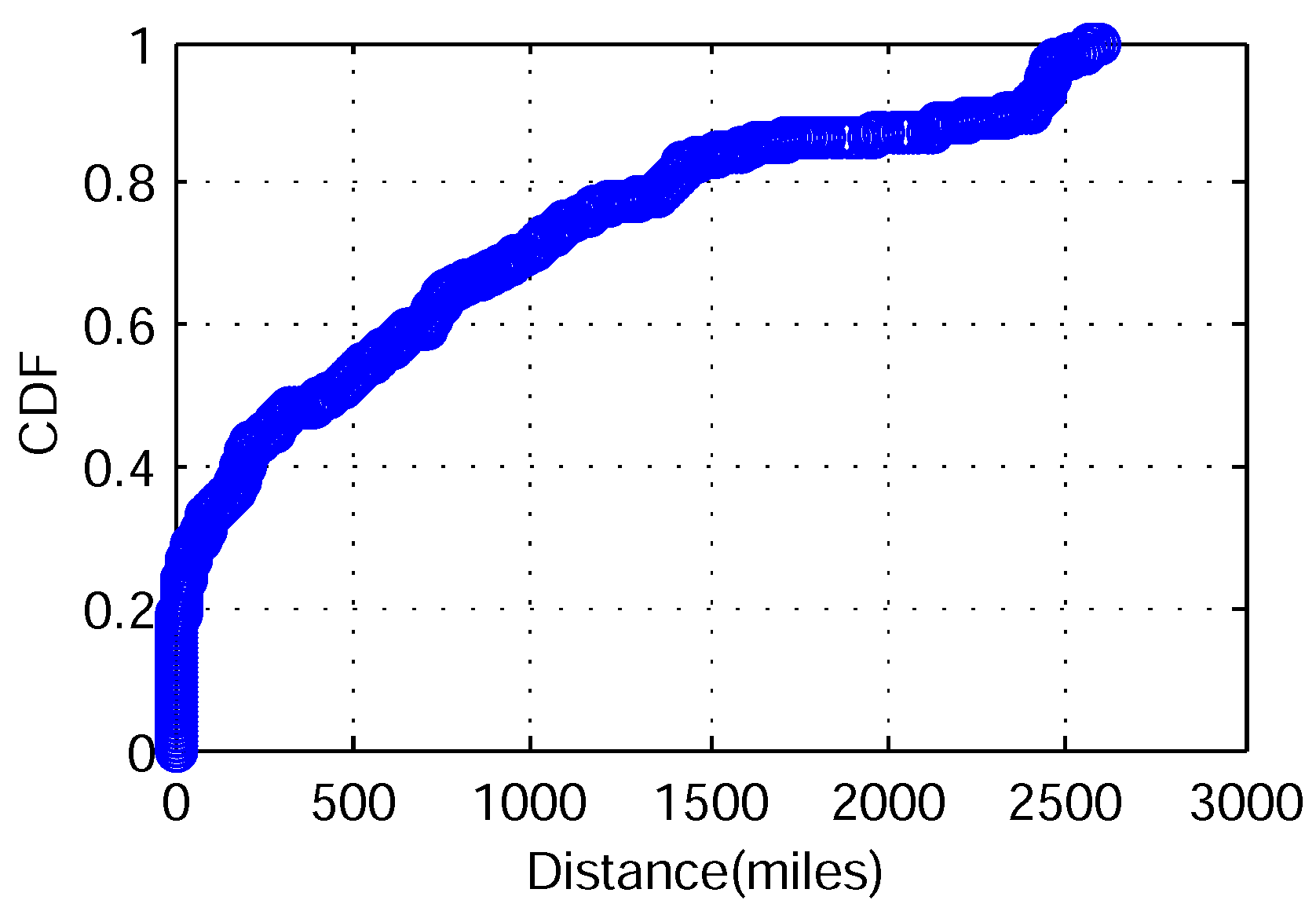}
	}
	\caption{CDFs of distance between a center point and each geotagged tweet in two cases.}
	\label{fig:SpatialCDF}
	\vspace{-15pt}
\end{figure}

To measure locality of Ebola-related tweets, we propose and use three locality measures based on previous studies \cite{brodersen2012youtube,kamath2013spatio}: (i) Ebola focus; (ii) Ebola entropy and (iii) Ebola spread.


Given a set of Ebola-related tweets/messages $M=\{m_1, m_2, m_3, ..., m_a\}$ and a set of locations $L=\{l_1, l_2, l_3, ..., l_b\}$, Ebola focus during a time interval (e.g., 30 minutes) is defined as follows:
\vspace{-5pt}
\begin{equation}
\label{eq:EbolaFocus}
EbolaFocus_{\bigtriangleup t}=\dfrac{1}{n_{\bigtriangleup t}} \max\limits_{l_i \in L} (n_{\bigtriangleup t}^{l_i})
\vspace{-5pt}
\end{equation}

, where $n_{\bigtriangleup t}$ is the total number of tweets posted during a time interval $\bigtriangleup t$ and $n_{\bigtriangleup t}^{l_i}$ is the number of tweets posted at a certain location $l_i$ during the time interval.

Equation~\ref{eq:EbolaFocus} is to measure the largest fraction of tweets posted in a location during the time interval. The larger Ebola focus is, the more number of tweets was posted in a single location/region. Intuitively, as Ebola-related tweets were propagated from one location/region to multiple locations/regions, Ebola focus will be reduced.

Next, we define Ebola entropy, which measures spatial distribution of Ebola-related tweets during a time interval, as follows:
\vspace{-5pt}
\begin{equation}
\label{eq:EbolaEntropy}
EbolaEntropy_{\bigtriangleup t}= -\sum_{l_i \in L} \dfrac{n_{\bigtriangleup t}^{l_i}}{n_{\bigtriangleup t}}     \log_2 \dfrac{n_{\bigtriangleup t}^{l_i}}{n_{\bigtriangleup t}}
\vspace{-5pt}			
\end{equation}

If all of Ebola-related tweets are posted in a single region/location, Ebola entropy will be zero. As tweets are posted in more number of locations/regions, Ebola entropy will increase accordingly, reflecting spread of Ebola information.

To measure how far Ebola-related information was spread, we define Ebola spread of Ebola-related tweets as:
\vspace{-5pt}
\begin{equation}
\label{eq:EbolaSpread}
EbolaSpread_{\bigtriangleup t} = \frac{1}{|L|} \sum_{i=1}^{|L|} D(l_0, l_i)
\vspace{-5pt}
\end{equation}

which measures the mean distance for all Ebola-related tweets of an Ebola case/event from its geographic midpoint $l_0$. The midpoint of the those tweets is measured by Haversine distance, taking into account the curvature of the Earth.

In the following sub-sections, we use Ebola focus, entropy and spread to understand how Ebola-related tweets were propagated.


	

\vspace{-5pt}
\subsection{Analysis of Spatial Properties}
\vspace{-5pt}

Given two events (i.e., Dallas and New York City cases), we analyzed spatial properties like how far Ebola-related tweets were spread, and what Ebola focus, entropy and spread values of the two events are. Figure~\ref{fig:SpatialCDF} shows CDFs of distance between a center point and each geotagged tweet of each event. We observed that only a small portion of tweets was spread locally. For example, about 7\% and 29\% of tweets were posted within 50 miles in Dallas and New York City cases, respectively. In contrast, most Ebola-related tweets were widely spread. For example, in Dallas case, about 46\% of tweets were posted in between 100 and 1,000 miles while about 41\% of tweets were posted in between 1,000 and 1,500 miles. In New York City case, 37\% of tweets were posted in between 100 and 1,000 miles while 13\% of tweets were posted in between 1,000 and 1,500 miles and 16\% of tweets were posted in over 1,500 miles from a center point. This result shows that both cases got attention from people nationally.

\begin{figure}
	\centering
	\subfigure[Dallas case] 
	{
		\label{fig:HalfHourlyDisDallas)}
		\includegraphics[width=0.46\linewidth,height=1.3in]{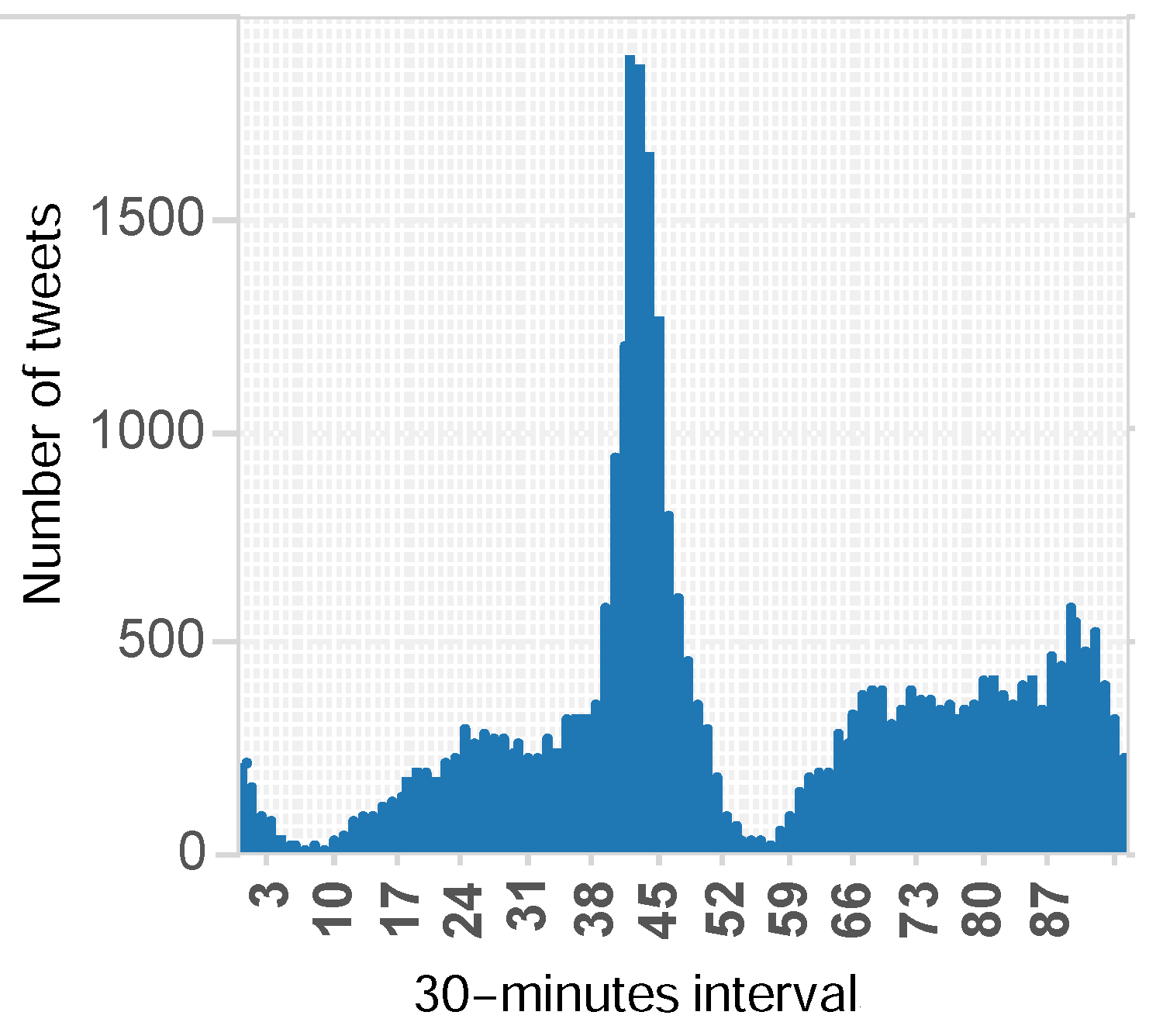}
	}
	\subfigure[New York City case] 
	{
		\label{fig:HalfHourlyDisNewYork}
		\includegraphics[width=0.46\linewidth,height=1.3in]{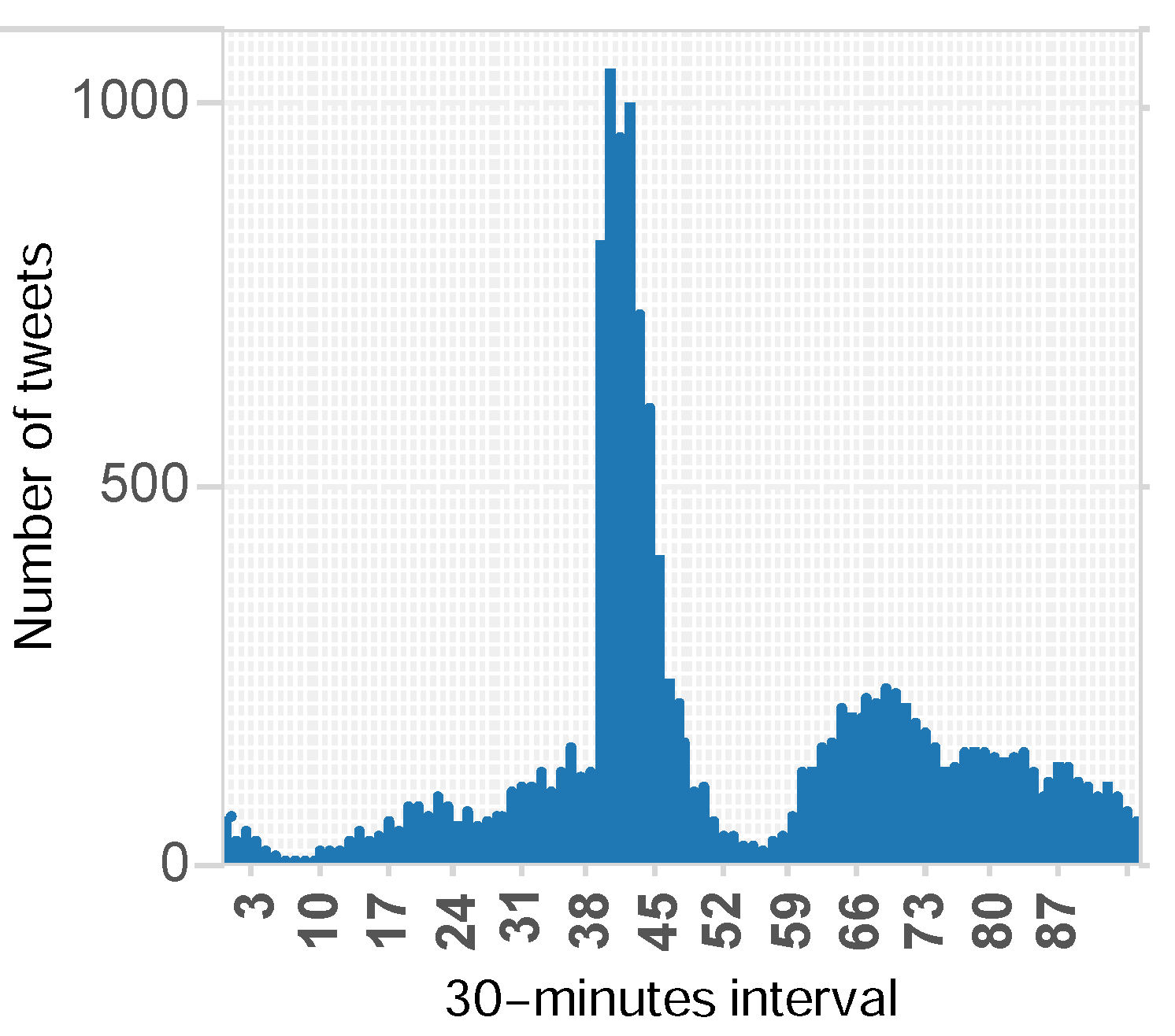}
	}
	\caption{Distributions of tweets in each 30-minute interval of two events.}
	\label{fig:HalfHourlyDis}
	\vspace{-10pt}
\end{figure}

\begin{figure*}
	\centering
	\subfigure[Ebola focus] 
	{
		\label{fig:FocusTemporalEvo-notCumulative}
		\includegraphics[width=0.31\linewidth]{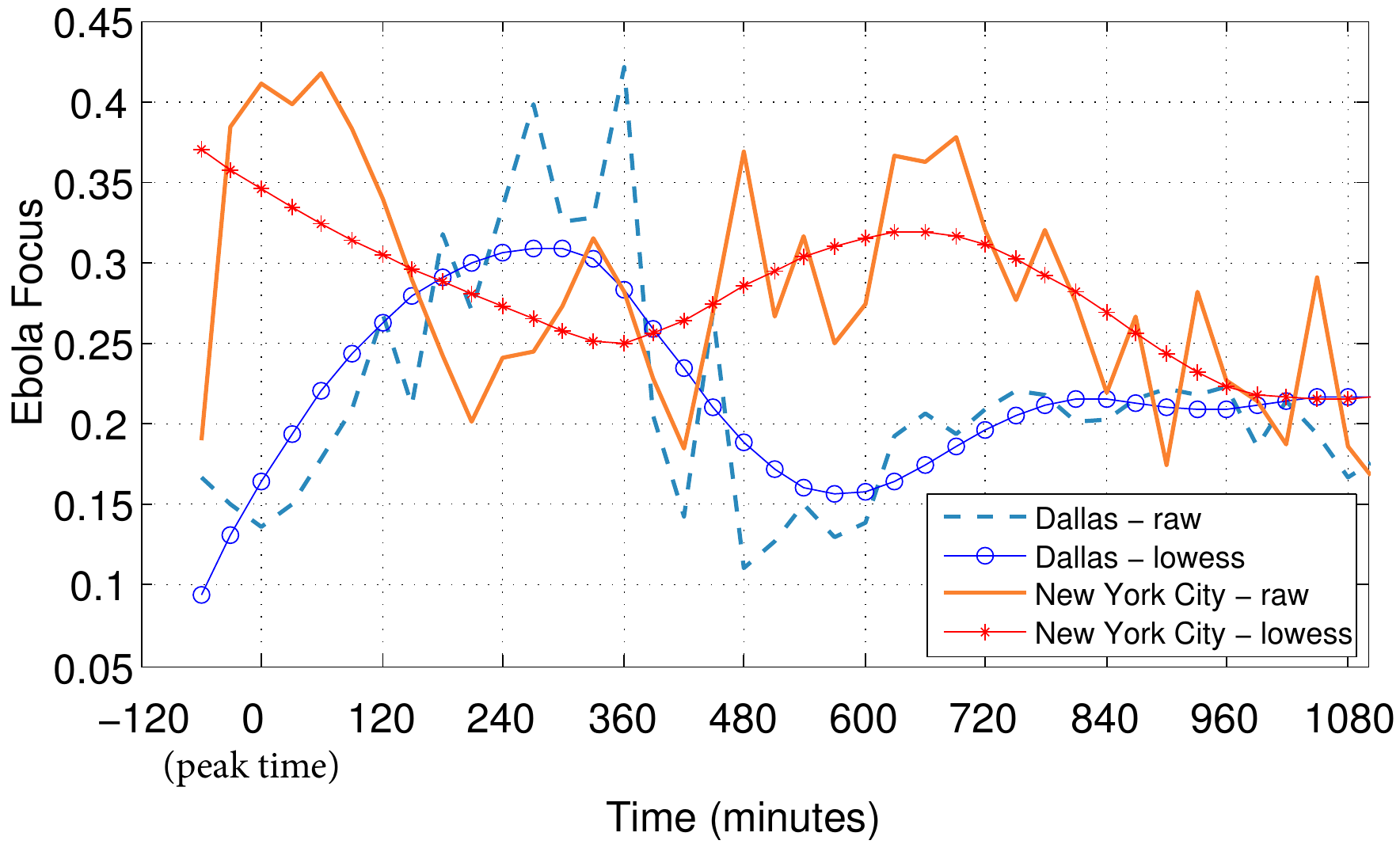}
	}
	\subfigure[Ebola entropy] 
	{
		\label{fig:EntropyTemporalEvo-notCumulative}
		\includegraphics[width=0.31\linewidth]{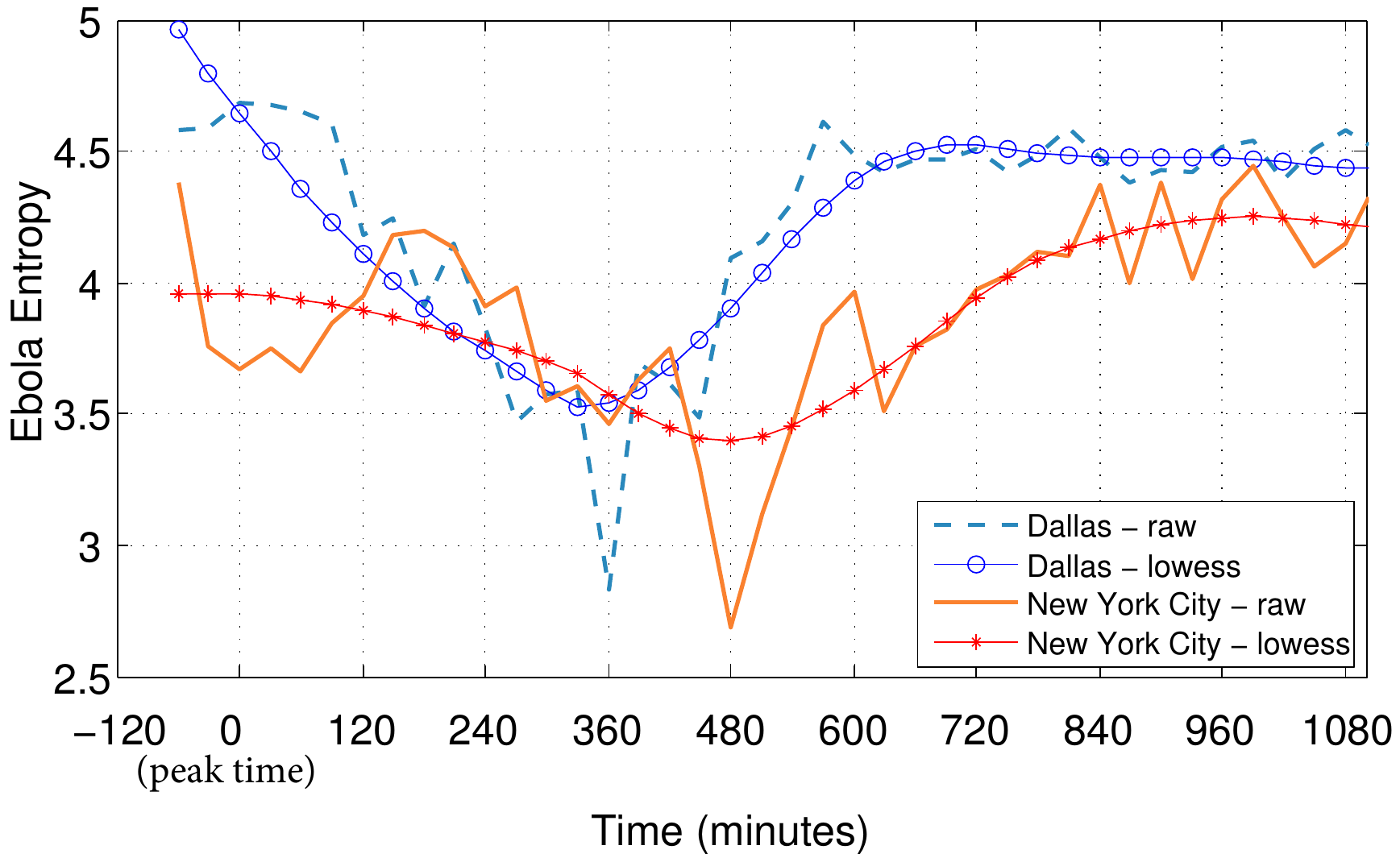}
	}
	\subfigure[Ebola spread] 
	{
		\label{fig:SpreadTemporalEvo-notCumulative}
		\includegraphics[width=0.31\linewidth]{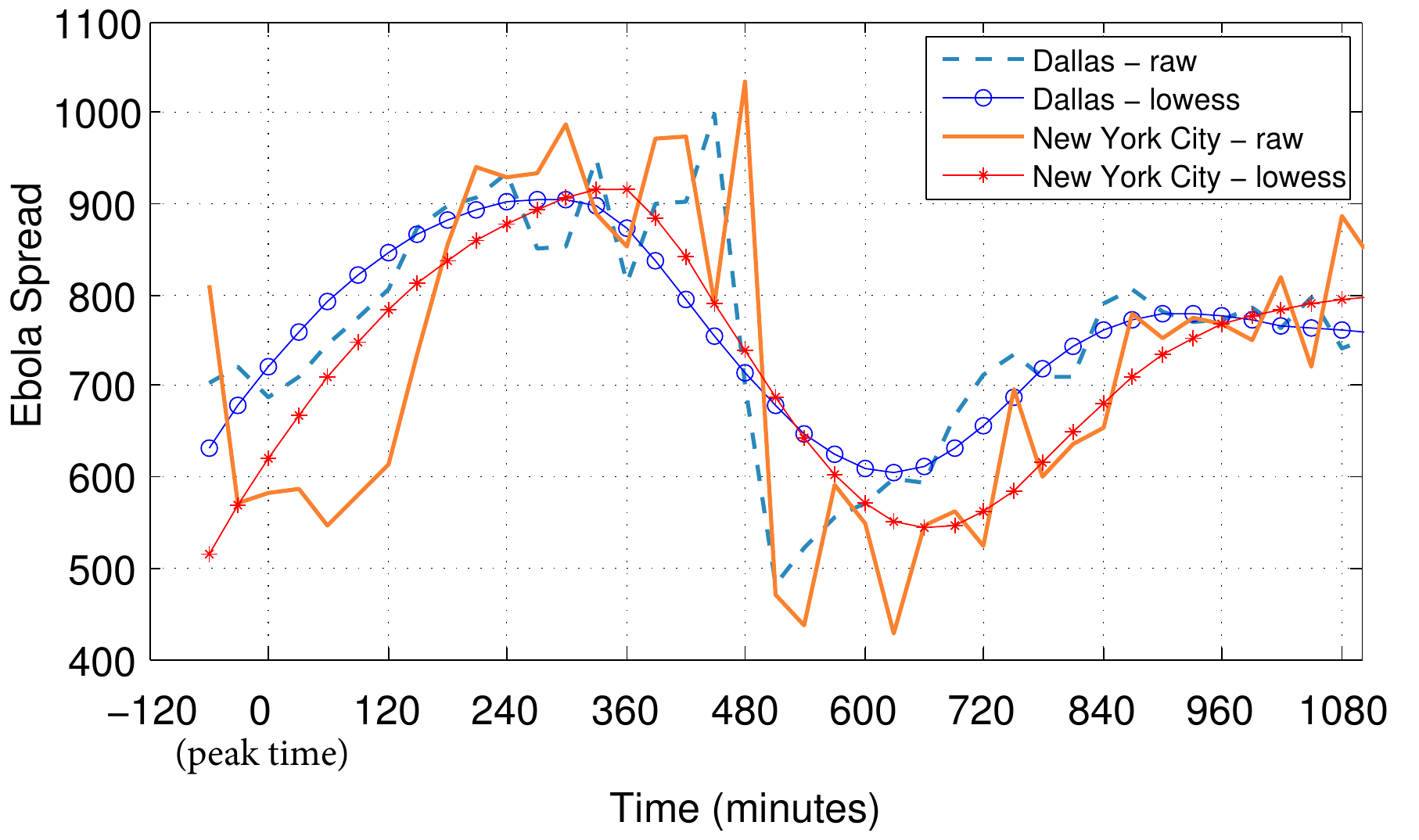}
	}
	\caption{Ebola focus, entropy and spread in 30-minute interval in Dallas and New York City cases.}
	\label{fig:TemporalImpact-notCumulative} 
	\vspace{-15pt}
\end{figure*}

\begin{table}
    \centering
    \caption{Ebola focus, entropy and spread of Dallas and New York City cases.}
    \scriptsize
    \small
	\begin{tabular}	{lrrr}
		\toprule		
		Ebola Case & Focus  & Entropy  & Spread (miles) \\
		\midrule
		Dallas 	& 0.19  & 4.62 & 774\\
		New York City 	& 0.27  & 4.27 & 734 \\
		\bottomrule
	\end{tabular}
	\label{table:FocusEntropyandSpread}
    \vspace{-15pt}
\end{table}

Next, we measured Ebola focus, entropy and spread for each of the two events. In particular, in each event, we grouped 46,598 tweets by a US state, considering each state as a location for Ebola focus and Ebola entropy in Equations~\ref{eq:EbolaFocus} and ~\ref{eq:EbolaEntropy}. For Ebola spread, we used Equation~\ref{eq:EbolaSpread}. As shown in Table~\ref{table:FocusEntropyandSpread}, focus, entropy and spread of Dallas case were 0.19, 4.62 and 774, respectively while focus, entropy and spread of New York City case were 0.27, 4.27 and 734, respectively. New York City case had larger focus, smaller entropy and smaller spread than Dallas case. It means tweets related to New York City case were posted more locally or in nearby areas than Dallas case.

\vspace{-5pt}
\subsection{Analysis of Temporal Properties}
\vspace{-5pt}


So far we analyzed spatial properties like how far Ebola-related tweets were propagated in US and measured Ebola focus, entropy and spread of the two events. Other interesting research questions are: (i) did both events have similar patterns in terms of the number of posted tweets over time?; and (ii) how did Ebola focus, entropy and spread of the events have been changed over time after a peak?


To answer the first question, we divided tweets in each event (again, two day period) by every 30 minutes. Figure~\ref{fig:HalfHourlyDis} shows distributions of tweets in each 30-minute interval for two events. 0 in \emph{x}-axis means 00:00 (midnight) of the first day. Tweets reached a peak at 42nd interval (9pm) in Dallas case and 40th interval (8pm) in New York City case. Overall, tweet distributions of the events were similar in the first day reaching the peak point, but in the second day users consistently posted tweets in Dallas case than New York City case. We conjecture that since Dallas case is the first imported Ebola case in US, people had longer attention and posted news and responses consistently.

To answer the second question, we measured focus, entropy and spread in each 30-minute interval as shown in Figure~\ref{fig:TemporalImpact-notCumulative}. 0 in \emph{x}-axis means the time when the number of tweets reached a peak (again, 9pm on October 1 in Dallas case and 8pm on October 23 in New York City case). Since there was some noisy up-and-down in the raw lines, we applied locally weighted scatter plot smoothing (LOWESS) \cite{cleveland1981lowess} to make the raw lines smoother. Then we analyzed the smoother lines (i.e., LOWESS lines). Focus in Dallas case gradually increased until 270 minutes, then decreased until 600 minutes. It next increased again until 840 minutes, and became stable. It means people nearby Dallas/Texas areas posted more tweets until 270 minutes by paying attention on the Ebola case. Then people outside Dallas/Texas posted more tweets between 270 and 600 minutes, by paying more attention on the case and related news and indicating Ebola-related tweet propagation. Entropy in Dallas case decreased until 360 minutes and then increased until 690 minutes with Ebola-related tweet propagation. Spread in Dallas case increased until 270 minutes after a peak, indicating that even though more tweets were locally posted (increasing focus), people in far areas (e.g., Washington state) also increased the number of related tweets. Then, spread decreased until 630 minutes, indicating people in far areas relatively decreased their tweets compared with local people. Then, spread gradually increased until 900 minutes and became stable.

Interestingly, New York City case had different patterns in the change of Ebola focus and entropy over time. In particular, focus in New York City case decreased until 360 minutes after a peak, reaching 0.25 focus. Then, focus increased until 660 minutes and then decreased. Entropy decreased until 480 minutes after a peak and then gradually increased. However, New York City case had similar spread pattern with Dallas case. Overall, New York City case had larger focus, smaller entropy and slightly smaller spread compared with Dallas case. It indicates that Ebola-related information in Dallas case was more widely propagated (or Dallas case got attention from geographically wider areas) since it was the first imported case in US.

\vspace{-5pt}
\subsection{Analysis of Social Properties}
\vspace{-5pt}
On Twitter, a followee's posting is automatically displayed to his follower's timeline. In Ebola cases, we are interested in how social ties play an important role in spreading Ebola-related tweets. In other words, will followers post/retweet more Ebola-related tweets as time goes after followees posted Ebola-related tweets?

To answer the question, we crawled Twitter following network of users and built propagation trees in which there are two types of nodes: (i) a root node; and (ii) a child node. A root node is a user who first posted an Ebola-related tweet in his social network (i.e., none of his followees had posted Ebola-related tweets before). A child node is a user who posted an Ebola-related tweet and at least one of his followees had posted an Ebola-related tweet before the user.

Next, we examined what percentage of nodes were child nodes in each 30-minute interval. We cumulatively added nodes/users, who posted Ebola-related tweets, as we added more 30 minutes. In the meantime, we measured how a proportion of child nodes changed over time.

\begin{figure}
	\centering
	\includegraphics[width=0.4\textwidth]{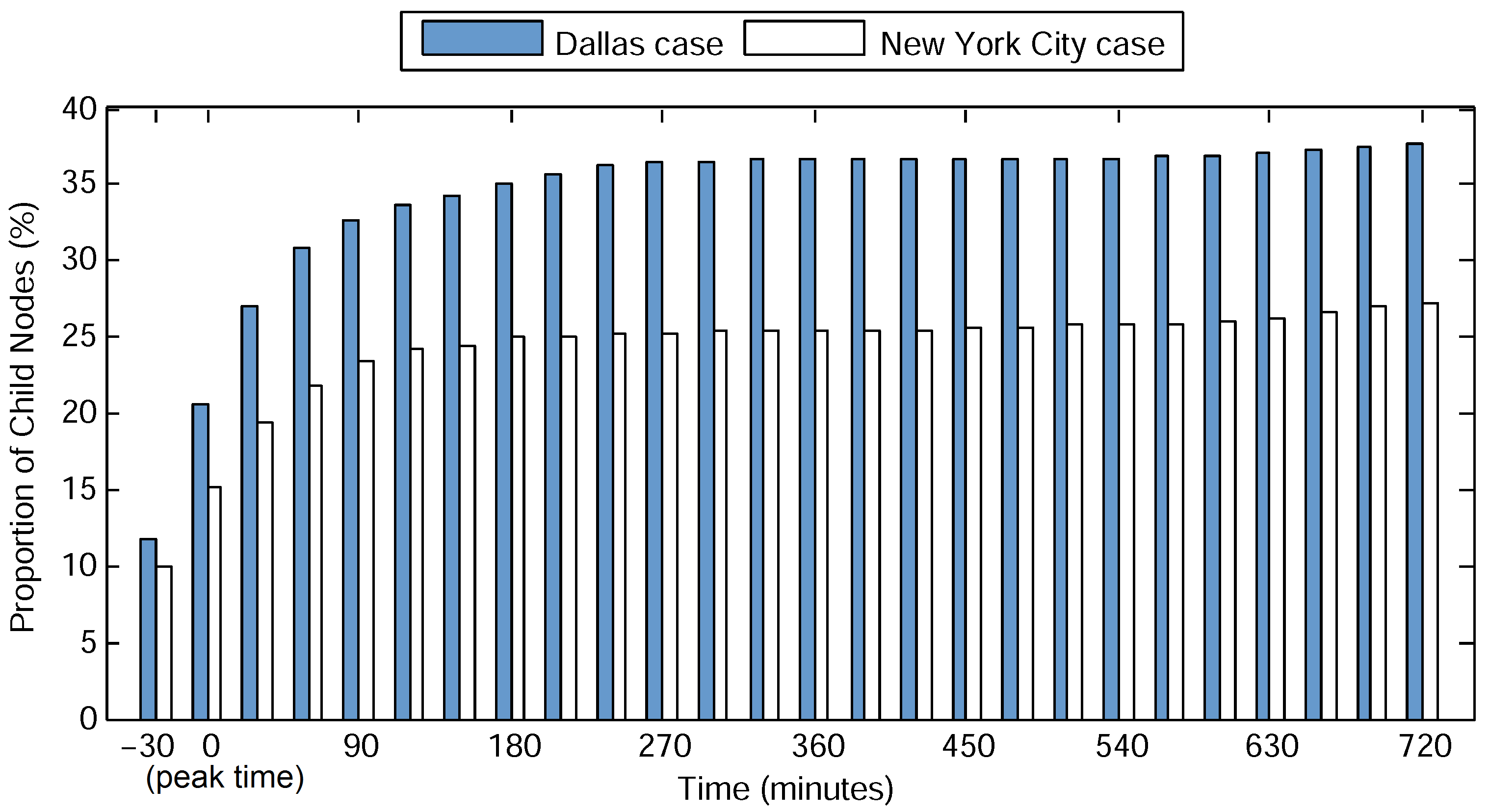}
	\caption{Change of a proportion of child nodes in different time slots of Dallas and New York City cases.}
	\label{fig:SocialImpactDallasNewYork}
\vspace{-15pt}
\end{figure}

Figure~\ref{fig:SocialImpactDallasNewYork} shows the change of a proportion of child nodes over time. Again 0 in \emph{x}-axis is a peak of each event in terms of the number of posted Ebola-related tweets. In Dallas case, a proportion of child nodes gradually increased until 240 minutes after a peak, and then became stable, reaching 35\% (i.e., 35\% of nodes were child nodes). Likewise, in New York City case, a proportion of child nodes gradually increased until 120 minutes after a peak, and then became stable, reaching 23\%. This analysis shown that social ties played an important role to propagate Ebola-related information within a few hours after a peak. In other words, social properties/ties affected the propagation of Ebola-related information over social media.
\section{Focus, Entropy and Spread of Six Topics}
\vspace{-5pt}

Now we apply three locality measures -- Ebola focus, entropy  and spread -- to 6 LDA topics that we found in Section~\ref{sec:topic}. Given a set of Ebola-related tweets associated with each topic, we examined which topic was more globally or locally discussed by applying the locality measures to each topic. As shown in Table~\ref{table:TopicFocusEntropy}, topic 2 ``Ebola outbreak in the world'' had the lowest focus, highest entropy and highest spread which mean the topic was discussed more globally than the rest. It makes sense because the topic itself is related to the world. However, topic 5 ``Jokes, swear and against jokes'' and topic 6 ``Impact of Ebola to daily life'' had the highest focus, lowest entropy and lowest spread which mean that the topics were discussed more locally than the rest.

\section{Conclusion}
\vspace{-5pt}


In this paper, we have presented a comprehensive analysis of the spread of Ebola-related information. In particular, we have conducted a large-scale data-driven analysis of 0.77 million geotagged tweets extracted from 2 billion tweets. By building topic models, we have found 6 topics. 
Then, we have analyzed how spatial, temporal and social properties were related to propagation of those tweets. We have found that (i) tweets related to New York City case were more locally propagated than tweets related to Dallas case; (ii) New York City case had consistently larger focus, smaller entropy and smaller spread over time; and (iii) a proportion of child nodes gradually increased until a few hours after a peak of each case, meaning that social ties played an important role in spreading Ebola-related tweets. 
Our work provides new insights into Ebola outbreak by understanding citizen reactions and topic/event-based information propagation, as well as providing a foundation for analysis and response of future public health crises. 

\begin{table}
    \centering
    \caption{Ebola focus, entropy and spread of 6 LDA topics.}
    \scriptsize
    \small
	\begin{tabular}	{crrr}
		\toprule
		Topic & Focus & Entropy & Spread (miles)\\
		\midrule
		1	& 		  0.748 	& 		  1.906		& 1,918.36\\
		2	& \textbf{0.577} 	& \textbf{2.865}	& \textbf{3,043.10}\\
		3	& 		  0.719 	& 		  1.945		& 2,025.27\\
		4	& 		  0.711 	& 		  2.162 	& 2,280.94\\
		5	& 		  0.761 	& 		  1.728		& 1,812.36\\
		6	& \textbf{0.762} 	& \textbf{1.695}	& \textbf{1,810.23}\\
		\bottomrule
	\end{tabular}
	\label{table:TopicFocusEntropy}
    \vspace{-15pt}
\end{table}

\section*{Acknowledgment}
\vspace{-5pt}
This work was supported in part by NSF grant CNS-1553035. Any opinions, findings and conclusions or recommendations expressed in this material are the author(s) and do not necessarily reflect those of the sponsors.

{\small	
\bibliographystyle{IEEEtran}
\bibliography{IEEEexample}
}


%
%
%

\end{document}